\documentclass[twocolumn,amsmath,ammsymb]{revtex4-1}
\usepackage{epsfig}
\usepackage{amsmath}
\usepackage{amssymb}
\usepackage{graphicx}
\usepackage{dcolumn}
\usepackage{soul}
\usepackage{bm}
\usepackage[usenames]{color}

\begin{document}

\preprint{}

\title{Emergence of Asymmetric Fermionic Orders in Interacting Birefringent Fermions}
\author{Yi-Xiang Wang$^{1,2}$}
\author{Fuxiang Li$^3$}
\affiliation{$^1$School of Science, Jiangnan University, Wuxi 214122, China.}
\affiliation{$^2$Department of Physics and Astronomy, University of Pittsburgh, Pittsburgh, PA 15260, USA}
\affiliation{$^3$School of Physics and Electronics, Hunan University, Changsha 410082, China}

\date{\today}

\begin{abstract}
The birefringent fermions possess a spectrum with two distinct Fermi velocities.  Here based on the lattice model, we use the mean-field method to investigate the interaction-induced phase transitions of the birefringent fermions.  We consider both the short-range nearest-neighbor (NN) and next-nearest-neighbor (NNN) repulsive interactions and calculate the phase diagrams under different conditions.  We find that the NN interactions can induce the asymmetric charge density wave order, while the NNN interactions can drive the asymmetric quantum anomalous Hall order (AQAH) in the large limit of anisotropy in the hopping integrals, $\beta\rightarrow1$.  The AQAH order is characterized by the unequal loop currents connecting the NNN sites and by the appearance of a gap between the two conduction (valence) bands.  Such asymmetric fermionic orders can be attributed to the specific lattice geometry of the birefringent fermions.  The implications of our results in experiments are also discussed.
\end{abstract}

\maketitle

\section{Introduction}

In condensed matter physics, the emergent quasiparticles in materials have close analogs in high energy physics and much progress has been made in the past few years.  The famous example is the behavior of two-dimensional (2D) Dirac fermions, which can be modeled by the Dirac equation, as in graphene \cite{A.H.Castro}, or on the surface of topological insulators \cite{M.Z.Hasan, X.L.Qi, A.Bansil}.  Other unconventional quasiparticles, such as Majorana fermions \cite{L.Fu} as well as  three-dimensional Weyl fermions \cite{N.P.Armitage}, are also solutions of the Dirac equation and have been detected in condensed matter systems.  The 2D birefringent fermions are massless fermions and differ from the common Dirac fermions in that they have two distinct velocities, rather than one, and therefore break the Lorentz symmetry.  They were proposed through the time-varying quadrupolar potential and time-varying hoppings in both $x-$ and $y-$directions \cite{M.P.Kennett} and may be implemented on the ultracold atoms in optical lattices \cite{A.S.Sorensen}.  With the help of the time-evolving operator, the model is reduced to a static effective one and can be represented in a particular tight-binding model on the square lattice, which is closely related to the Lieb lattice \cite{C.Weeks, W.-F.Tsai, V.I.Iglovikov}.  The birefringent fermions can be compared to the high-spin, such as spin-$\frac{3}{2}$ Dirac fermions \cite{B.Roy2018}.  But the velocities of spin-$\frac{3}{2}$ Dirac fermions are fixed to the values proportional to $\frac{3}{2}$ and $\frac{1}{2}$, while the birefringent fermions can continuously tune the velocities of the Dirac cones by controlling the parameter $\beta$ in experiments.

The properties of birefringent fermions have aroused many interests.  Their responses to a variety of perturbations \cite{M.P.Kennett} have been studied, as well as the topological defects \cite{B.Roy2012}, and it was found that the birefringent fermions exhibit certain robustness.  An important question is how the birefringent fermions will response to the interactions.  A fundamental model for describing the physics of interacting fermions on a lattice is the Hubbard model, including the kinetic energy $t$ between neighboring lattice sites and the interactions $U$ of opposite spins on the same site.  The Hubbard model may also be extended to the spinless case and include the short-range interactions, such as the nearest-neighbor (NN) and next-nearest-neighbor (NNN) ones.  Based on the Hubbard model, the interactions between fermions can drive the formation of various long-range orders, with their properties depending crucially on the combined effects of kinetic energy, interactions, lattice geometry and even dimension \cite{T.I.Vanhala, D.Prychynenko, A.M.Cook, W.Zheng, V.S.Arun, K.Jiang, Y.C.Zhang, B.Roy2017, Y.X.Wang2017, S.W.Kim, Y.X.Wang2018, Y.X.Wang2019}.

In this paper, we try to study the problem of the stability of spinless birefringent fermions to the NN and NNN interactions.  Similar problem has been studied in a previous work, assuming the linear dispersion all the way out to the large momentum \cite{N.Komeilizadeh}.  It was found that there exist the charge density wave (CDW) order and the quantum anomalous Hall (QAH) order driven by the short-range interactions.  Here our study is based on the lattice model, and thus the full spectrum can be taken into account, not only including the linear spectrum, but also the nonlinear spectrum as the momentum increases. 

By using the standard mean-field method to decouple the many-body interactions, the interacting phase diagrams are numerically calculated and identified under different conditions, and the phenomena of the interaction-driven metal-insulator transitions are explored.  The main results are as follows: (i) An analytical expression for the density of states (DOS) in noninteracting birefringent fermion system is derived, which may help us understand why the low-energy linear dispersions are incomplete in describing birefringent fermion, especially when the interactions set in the system.  (ii) The NN interactions can induce the asymmetric CDW (ACDW) order.  More interestingly, the NNN interactions can drive the asymmetric QAH (AQAH) order in the large limit of anisotropy in the hopping integrals, $\beta\rightarrow1$.  In comparison with the equal loop currents in the symmetric QAH (SQAH) order, the AQAH order is characterized by the unequal loop currents circulating around the NNN sites, and by the nonzero gap between the two conduction (valence) bands.  The emergence of such asymmetric fermionic orders can be attributed to the specific lattice geometry of the birefringent fermions. (iii) The interacting phase diagram of $\beta<0.8$ shows certain universality while that of $\beta>0.8$ includes a coexisting phase that incorporates both the ACDW and AQAH orders.  Our work may provide an important step forward in the ongoing effort to design the quantum materials with tailored properties.

\section{Model}

\begin{figure}
	\includegraphics[width=8.8cm]{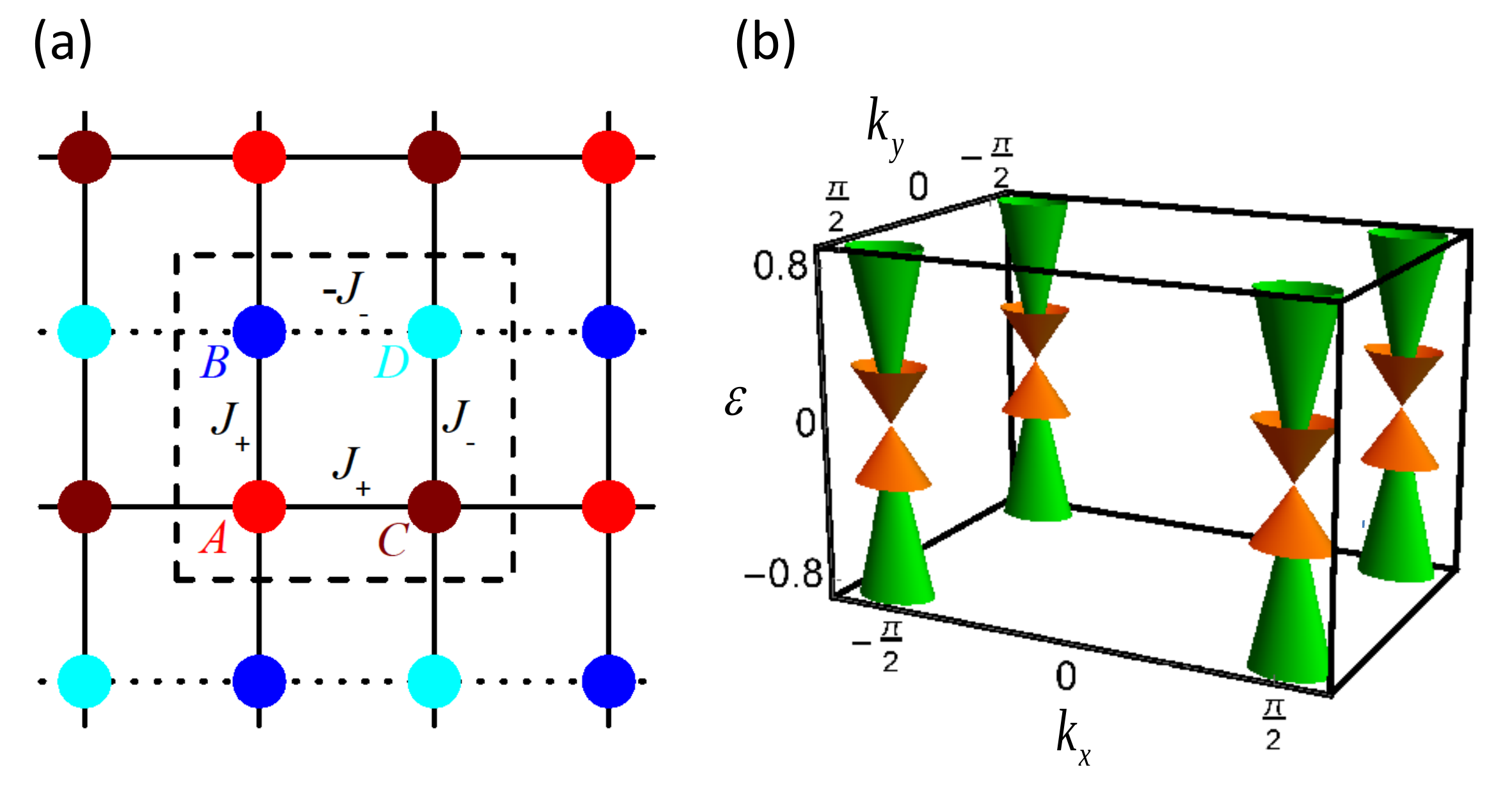}
	\caption{(Color online) (a) Schematic plot of the effective tight-binding model, with the unit cell and the hopping parameters being indicated.  The solid (dotted) lines represent the positive (negative) hoppings.  (b) The low-energy dispersions $\varepsilon(k_x,k_y)$ in the 2D Brillouin zone are shown when $\beta=0.5$, where the green/orange cones represent the $J_{+/-}$ bands, respectively. }
	\label{model}
\end{figure}

The spinless birefringent fermions were proposed to be related to the artificial magnetic field.  The effective model is constructed on a square lattice and is schematically plotted in Fig.~\ref{model}(a), with the spatially periodic magnetic field as well as the spatially periodic hopping amplitudes.  The unit cell includes four sublattices $A$, $B$, $C$ and $D$, and each plaquette is threaded by half a flux quantum $\frac{1}{2}\Phi_0$.  The noninteracting Hamiltonian is written as: 
\begin{align}
\hat H_0=\sum_{\langle ij\rangle} (J_{ij}e^{i\varphi_{ij}}\hat c_i^\dagger\hat c_j
+\text{H.c.}),
\end{align}
here $\hat c_i$ and $\hat c_i^\dagger$ denote the fermionic annihilation and creation operators at site $i$, respectively.  $J_{ij}=J_\pm=J_0(1\pm \beta)$ is the hopping integral between the NN sites $i$ and $j$, with $\beta\in[0,1]$ characterizing the anisotropy in the hopping integrals.  $\varphi_{ij}$ is the Peierls phase caused by the magnetic flux, which in Landau gauge can be given as $\varphi_{i,i+\hat x}=\pi i_y$ and $\varphi_{i,i+\hat y}=0$.  In the primitive proposal~\cite{M.P.Kennett}, the controlling parameter $\beta$ is considered as the hopping in $y-$direction relative to that in $x-$direction and can be tuned in experiment. 

The dispersions for $H_0$ are given as
\begin{align}\label{dispersion}
\varepsilon_{\eta \pm}(\boldsymbol k)=2\eta J_\pm \sqrt{d(\boldsymbol k)},
\end{align}
with $\eta=\pm$ being the conduction/valence band and $d(\boldsymbol k)=\text{cos}^2k_x+\text{cos}^2k_y$.  As the conduction and valence bands of both $J_{+/-}$ branch touch at zero energy (see Fig.~\ref{model}(b)), the noninteracting model supports the birefringent semimetal (BRS) phase.  When $\beta=0$, the model owns two copies of Dirac fermions.  In the limiting case of $\beta=1$, the model accommodates a Dirac cone and two completely flat bands at zero energy, as for the Lieb lattice \cite{M.R.Slot, R.Drost}.  In the following, we will set $J_0=1$ as the unit of energy.  From the dispersions, we can see that four equivalent Dirac points $\boldsymbol K_{\pm\pm}=(\pm\frac{\pi}{2}, \pm\frac{\pi}{2})$ are located at the corners of the first Brillouin zone, as shown in Fig.~\ref{model}(b).  Around $\boldsymbol K_{++}$, the low-energy dispersions are expanded as
\begin{align}\label{low-en}
\varepsilon_{\eta \pm}(\boldsymbol k)=2\eta J_\pm \sqrt{k_x^2+k_y^2},
\end{align}
in which the linear dispersion $\varepsilon\sim|k|$ can be seen.

\begin{figure}
\includegraphics[width=9.2cm]{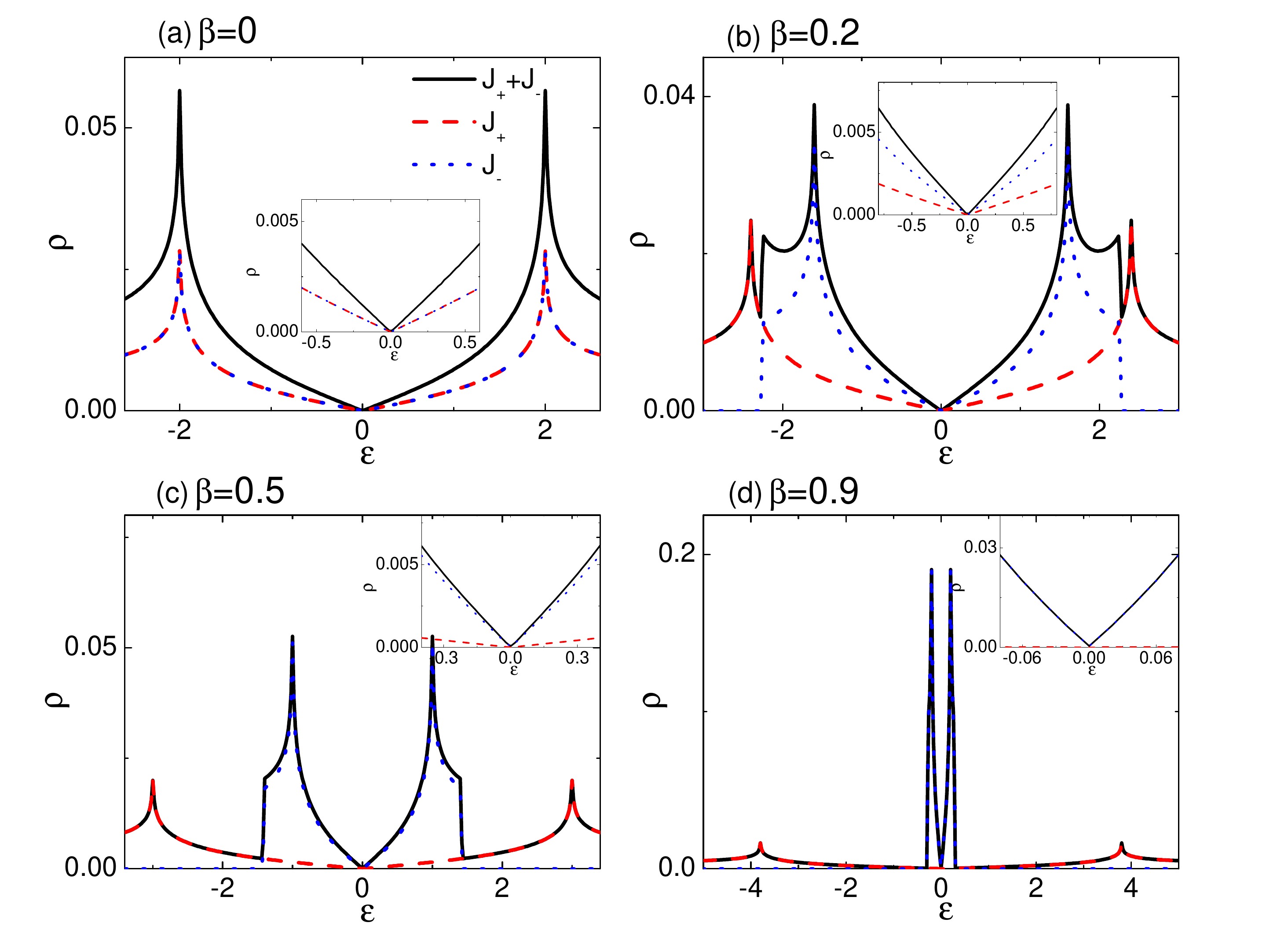}
\caption{(Color online) The evolution of the DOS of birefringent fermions with $\beta$, where the contributions from $J_+$, $J_-$ and both bands are indicated.  The insets are the expanded view of the DOS around zero energy.  The legends are the same in all figures. }
\label{DOS}
\end{figure}

For the DOS per unit cell, an analytical expression is derived and has the form
\begin{align}
&\rho(\varepsilon)=\rho_++\rho_-,
\\
&\rho_\pm=\frac{\theta(2\sqrt 2 J_\pm-|\varepsilon|)|\varepsilon|}{\pi^2J_\pm^2}
F\Big(\frac{\pi}{2},1-(\frac{\varepsilon^2}{4J_\pm^2}-1)^2\Big),
\end{align}
here $\rho_+$ and $\rho_-$ are the contributions from $J_+$ and $J_-$ band, respectively.  $F(\frac{\pi}{2},x)$ is the complete elliptic integral of the first kind and $\theta(x)$ is the step function.  As $F(\frac{\pi}{2},x)\simeq\frac{\pi}{2}$ when $x$ is around zero, we can get the low-energy expansion for the DOS as
\begin{align}
\rho(\varepsilon)=\frac{|\varepsilon|}{2\pi}(\frac{1}{J_+^2}+\frac{1}{J_-^2}),
\end{align}
where the linear DOS gives the characteristic of the Dirac fermions.

As shown in Fig.~\ref{DOS}, the DOS of the birefringent fermions shows interesting features as one varies the anisotropic parameter $\beta$. The linear DOS around zero energy is displayed in the insets.  When $\beta=0$ in Fig.~\ref{DOS}(a), $J_+$ and $J_-$ bands overlap and the Van-Hove singularities  lie at $|\varepsilon|=2J_0$ where the DOS diverges.  With the increasing of $\beta$, the Van-Hove singularities split and appear at $|\varepsilon|=2J_{+/-}$.   When $\beta>0.5$, if $|\varepsilon|<2\sqrt2J_-$, the DOS is dominated by $J_-$ bands.  When $\beta=0.9$ in Fig.~\ref{DOS}(d), the Van-Hove singularities of $J_-$ bands are moving to zero energy, and the linear low-energy region becomes much narrower, which is due to the asymptotic flatness of the outer $J_-$ bands when $\beta\rightarrow1$.  These interesting features in DOS provide a platform for the study of interaction-induced phases transitions in birefringent fermionic system. Indeed, as will be shown below, by taking into account of the full spectrum obtained from the lattice model, the fermionic orders of ACDW and AQAH will appear.

\section{Mean-field theory}

When the system is subjected to the short-range interactions, we consider the NN and NNN ones, which are given as
\begin{align}
\hat H_I=\hat H_U+\hat H_V,
\end{align}
with
\begin{align}
\hat H_U=U\sum_{\langle i,j\rangle}\hat n_i\hat n_j,
\qquad
\hat H_V=V\sum_{\langle\langle i,j\rangle\rangle}\hat n_i \hat n_j,
\end{align}
here $U,V>0$ denote the repulsive interaction strength between NN and NNN sites, respectively, and $\hat n_i=\hat c_i^\dagger\hat c_i$ is the fermionic number operator.

In the previous work~\cite{N.Komeilizadeh}, the authors assumed {\it a priori} the specific fermionic orders, and then performed the mean-field analysis based on the linear energy spectrum.  Here we consider the finite-size lattice model, which enables us to take into account all the properties of  the full spectrum and to explore new possible phases.  We do not assume any kind of fermionic orders at the beginning, but only decouple the two-body interaction operators into the possible Hatree and Fock channels within the mean-field strategy,
\begin{align}\label{decouple}
\hat c_i^\dagger \hat c_i \hat c_j^\dagger \hat c_j
\simeq&\langle \hat c_i^\dagger \hat c_i \rangle \hat c_j^\dagger \hat c_j
+\langle \hat c_j^\dagger \hat c_j\rangle \hat c_i^\dagger \hat c_i
\nonumber\\
&-\langle \hat c_i^\dagger \hat c_j \rangle \hat c_j^\dagger \hat c_i
-\langle \hat c_j^\dagger \hat c_i\rangle \hat c_i^\dagger \hat c_j
+\text{const}.
\end{align}
where $\langle\cdots\rangle$ denotes the average taken at the ground state and can be calculated self-consistently.  The first two terms are the Hartree terms while the third and forth are the Fock terms.  The last constant term will not affect the properties of the system, but must be included when calculating the total energy, as to determine the ground state.  Because the mean-field method can capture the different correlations when the parameters vary in a many-body system, it is qualitatively effective and reliable in dealing with the correlated systems, as has been demonstrated in Refs. \cite{A.M.Cook, W.Zheng, V.S.Arun, Y.X.Wang2017, Y.C.Zhang, K.Jiang, Y.X.Wang2018, Y.X.Wang2019, N.Komeilizadeh, S.W.Kim, B.Roy2017}. 

The self-consistent iterative steps are as follows \cite{V.S.Arun, D.Prychynenko, Y.X.Wang2017}: (i) set the initial random values for all terms in the brackets in Eq.~(\ref{decouple}), (ii) diagonalize the decoupled one-body Hamiltonian as to solve the energies and eigenvectors, (iii) use the obtained energies and eigenvectors to recalculate the terms in the brackets.  We take the convergence precision for two consecutive calculations to be $10^{-6}$.  If the convergence precision is reached, the iterative processes are completed, while if not, repeat the steps from (i) to (iii).  We then seek for the possible long-range fermionic orders based on the obtained results.  The self-consistent procedures may lead to a local minimum in energy.  To avoid this, we need to try several random configurations as the initial inputs, to help the mean-field procedure locate the ground state corresponding to the global minimum in energy.

As we focus on the bulk physics of the model, the calculations are performed  on the finite-size lattice with periodic boundary conditions.  Unless specified, we take the system size as $N_x=N_y=N_c=32$.  We consider the half-filling case, \textit{i.e.}, the number of fermions is half of the number of the lattice sites.

\section{Main Results}

\subsection{Only Nearest-neighbor Interactions}

\begin{figure}
	\includegraphics[width=8.8cm]{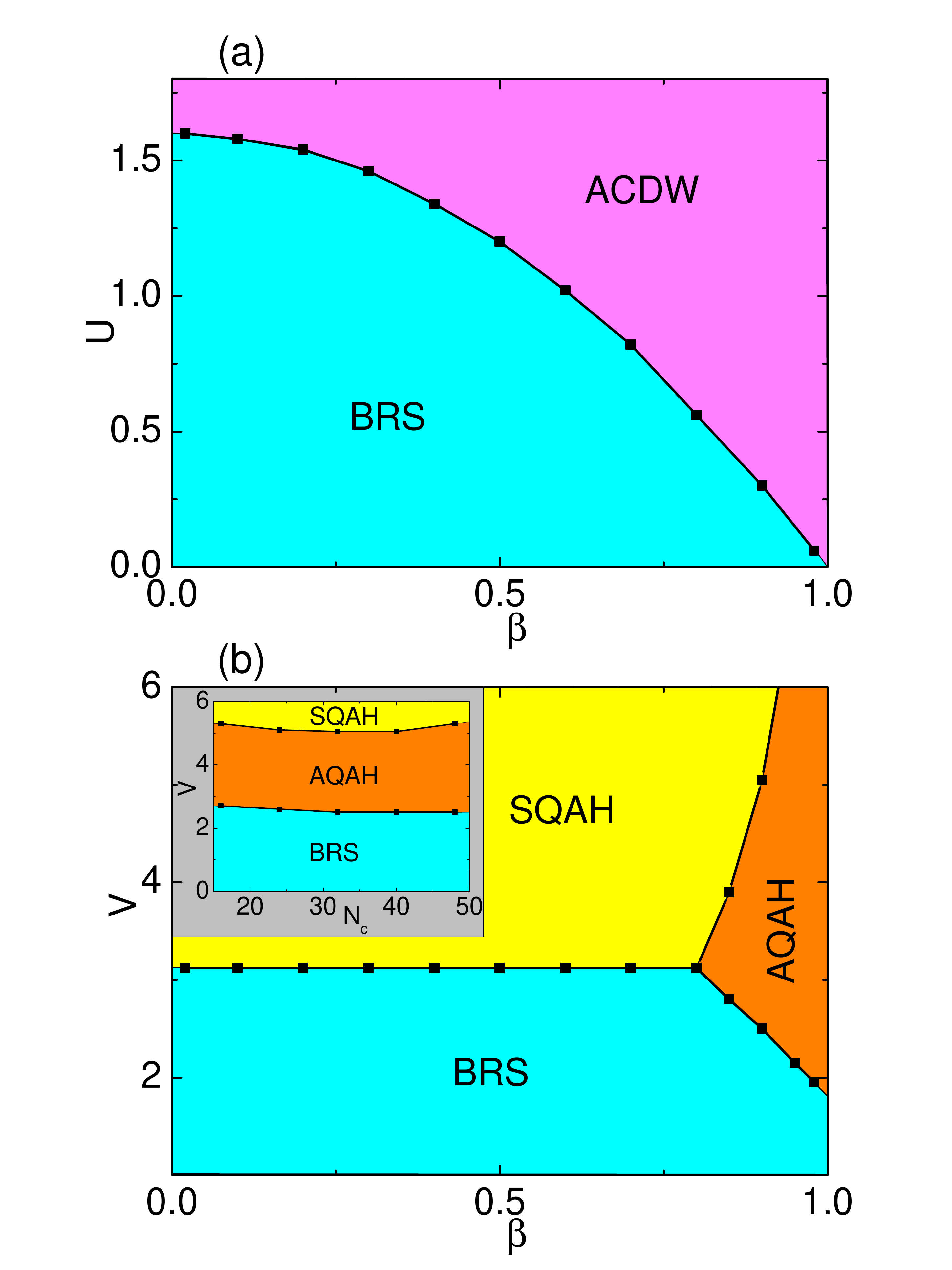}
	\caption{(Color online) Phase diagrams of interacting birefringent fermions in the parametric space of $\beta$ and interaction strength, with the presence of only NN interactions $U$ in (a) and only NNN interactions $V$ in (b).  The inset in (b) gives the dependence of AQAH on the lattice size $N_c$ when $\beta=0.9$. }
	\label{onlyUV}
\end{figure}

First we consider when only the NN repulsive interactions $U$ are present in the system.  The phase diagram in the parametric space $(\beta,U)$ is given in Fig.~\ref{onlyUV}(a), including two phases of BRS and ACDW.  In BRS phase, the fermions are equally distributed on each sublattice and the chiral symmetry that exchanges the neighboring sites is preserved.  We can see that the BRS phase is stable to weak $U$ due to the vanishing DOS at the Dirac points.  When $U$ increases to beyond the critical value, the fermion number $\langle \hat n_i\rangle$ on each site begins to fluctuate and breaks the chiral symmetry.  Then the spontaneous-symmetry-breaking CDW order dominates the system with the order parameter 
\begin{align}
Q=(\langle \hat n_A\rangle+\langle \hat n_D\rangle)-(\langle \hat n_B\rangle+\langle \hat n_C\rangle), 
\end{align}
and the corresponding phase transitions are continuous.  Note that the terms of $\langle \hat c_i^\dagger \hat c_j\rangle$ are always vanishing when only the NN interactions $U$ are included, meaning that the $U$ cannot drive the bond orders between the NN sites, but can only modulate the fermion number on each site. 

In Fig.~\ref{onlyUV}(a), we can see when $\beta$ increases, the critical $U_c$ decreases, thus the larger anisotropy in the hopping integrals favors the CDW order.  In the limit of $\beta=1$, $U_c$ tends to be vanishing.  This is because the sublattice $D$ is completely depleted in this case, which makes the unequal fermion number, or $Q\neq0$, more susceptible to the NN interactions.  That is, a tiny $U$ can drive the system into CDW.

\begin{figure}
	\includegraphics[width=9cm]{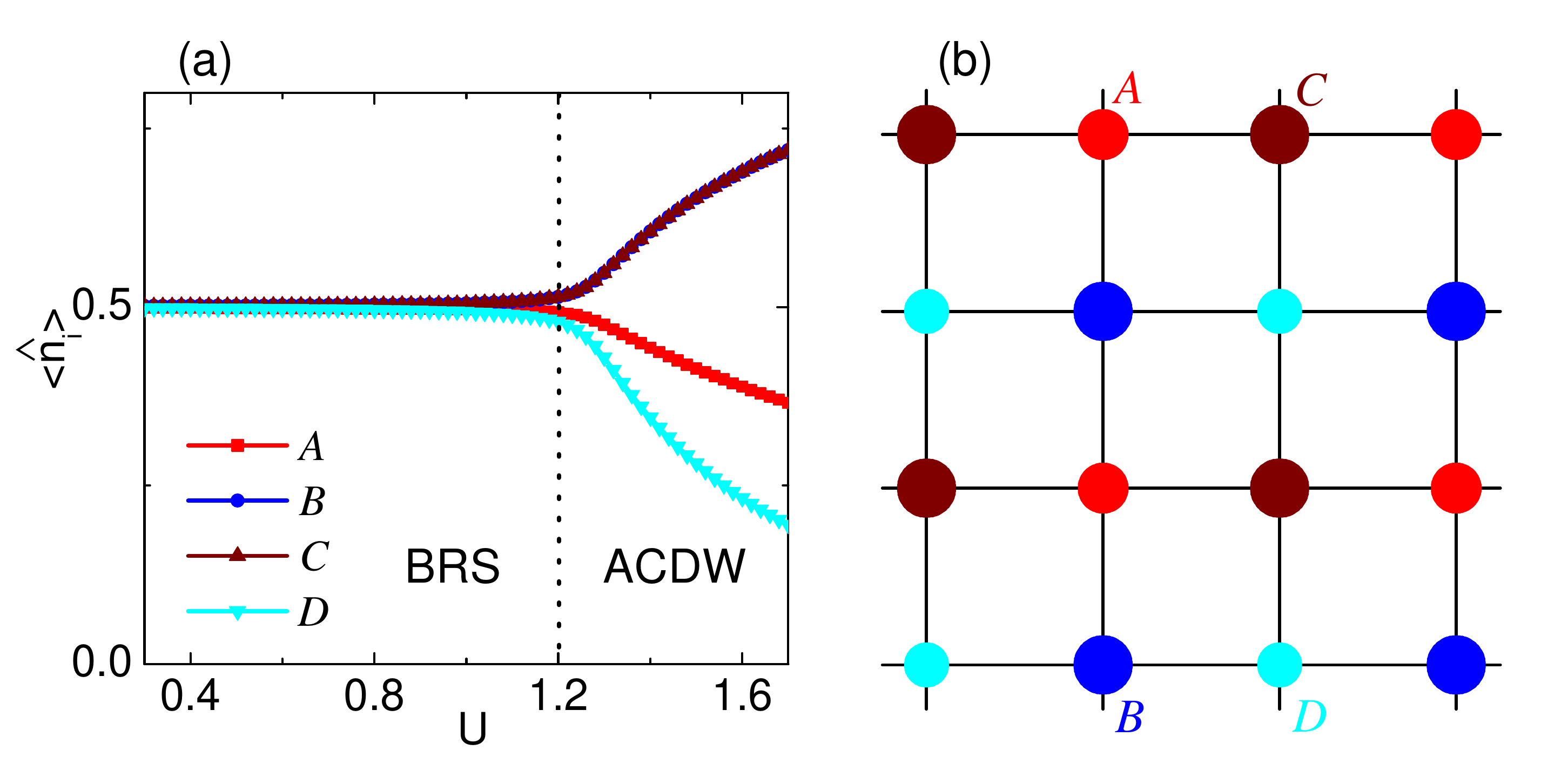}
	\caption{(Color online) Plot of the fermion number $\langle \hat n_i\rangle$ as a function of $U$ in (a) and the ACDW order in (b) driven by the NN interactions.  In (b) when $U=1.32$, the numerical results are $\langle \hat n_A\rangle=0.443$, $\langle \hat n_D\rangle=0.345$, $\langle \hat n_B\rangle=\langle\hat n_C\rangle=0.606$ and the size of the circle is proportional to the fermion number.  We choose the parameter $\beta=0.5$.}
	\label{CDW_order}
\end{figure}

The numerical results show that in the CDW order, the fermion number $\langle\hat n_B\rangle=\langle\hat n_C\rangle$ and $\langle\hat n_A\rangle\neq\langle\hat n_D\rangle$, so the NN interaction-induced CDW order is asymmetric.  In Fig.~\ref{CDW_order}(a) when $\beta=0.5$, we plot $\langle \hat n_i\rangle$ as a function of $U$, where the asymmetric fermion numbers can be clearly seen.  This can be explained by the equivalence between the sublattices $B$ and $C$ and the unequivalence between the sublattices $A$ and $D$.  Thus we have $\langle \hat n_B\rangle=\langle \hat n_C\rangle=\frac{1}{2}-\frac{Q}{4}$, $\langle \hat n_A\rangle=\frac{1+\delta n}{2}+\frac{Q}{4}$, $\langle \hat n_D\rangle=\frac{1-\delta n}{2}+\frac{Q}{4}$, with $\delta n$ being the fermion number difference between sublattices $A$ and $D$.  Note that $\delta n$ increases with $U$ in Fig.~\ref{CDW_order}(a).  The ACDW order is schematically plotted in Fig.~\ref{CDW_order}(b) with the chosen parameters, where the size of the circle is proportional to the fermion number.  Within the ACDW order, the dispersions are given as
\begin{align}
\varepsilon^U_{\eta\pm}(\boldsymbol{k})
=\eta\Big(4J_\pm^2 d(\boldsymbol k)+\frac{U^2Q^2}{4}\Big)^\frac{1}{2}, 
\end{align}
with $d(\boldsymbol k)$ being the same as Eq.~(\ref{dispersion}).  Clearly the ACDW order preserves the birefringent property, but can open a gap in the system.  Note that the fermion number difference $\delta n$ does not enter the dispersion.  

\subsection{Only Next-nearest-neighbor Interactions}

Next we consider only the NNN interactions $V$, with the phase diagram in the parametric space $(\beta,V)$ being given in Fig.~\ref{onlyUV}(b).  It shows that BRS phase is also robust to weak $V$.  When $V$ is strong, the spontaneous-symmetry-breaking QAH phase dominates and breaks the time-reversal symmetry of the system.  The related order parameter is defined as \cite{S.Raghu, K.Sun1}
\begin{align}
\Phi_{ij}=\langle\hat c_i^\dagger \hat c_j\rangle, 
\end{align}
here the sites $i$ and $j$ are connected by the NNN vectors.  Note that the terms of $\langle\hat c_i^\dagger \hat c_i\rangle$ are always equal to $\frac{1}{2}$ when the NNN interactions $V$ are included, meaning that $V$ cannot change the fermion number on each site. 

\begin{figure}
	\includegraphics[width=9cm]{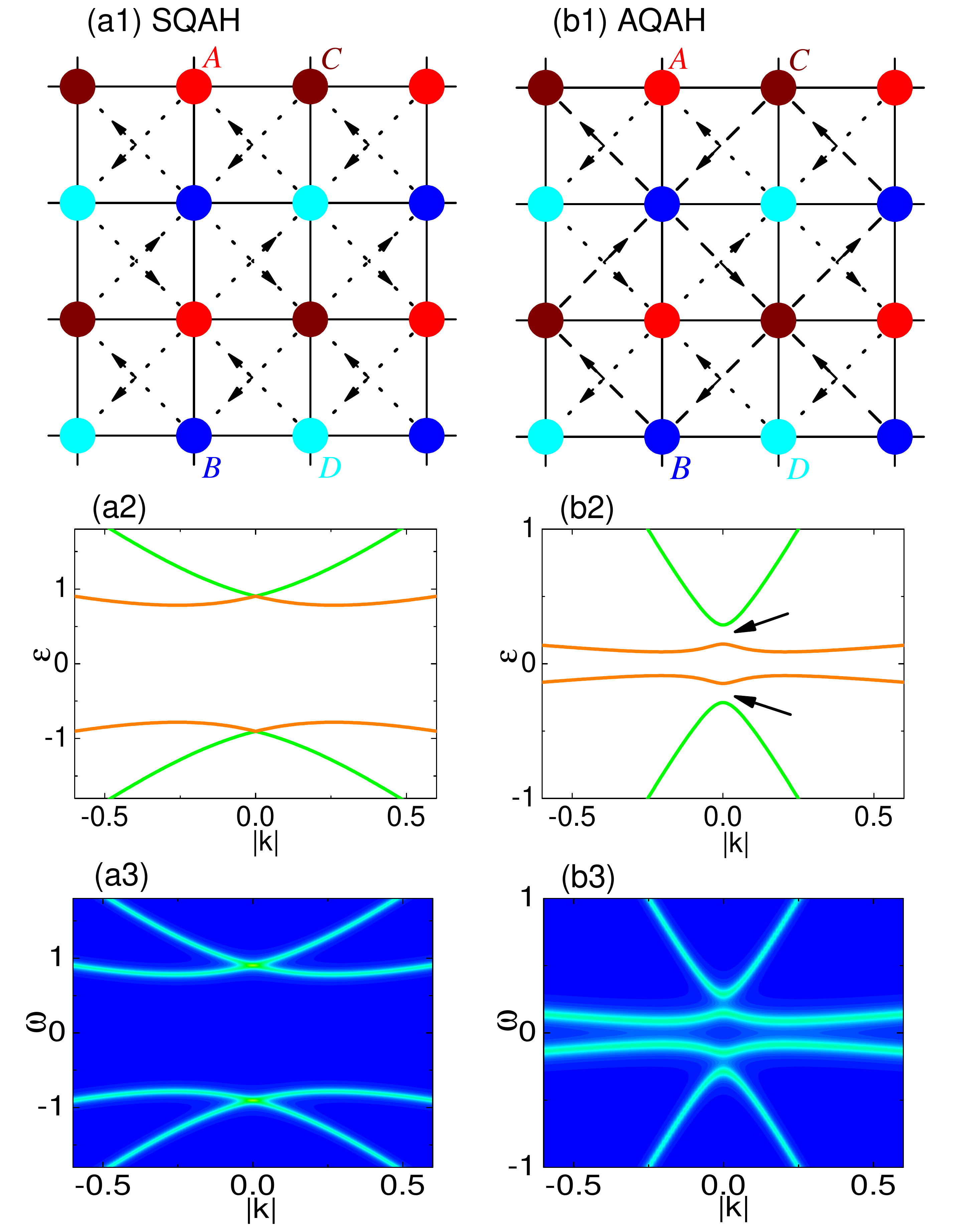}
	\caption{(Color online) Plot of the NNN hopping integrals (a1)-(b1), dispersions (a2)-(b2) and quasiparticle spectral density (a3)-(b3) due to the SQAH order and AQAH order driven by the NNN interactions.  In (a1) and (b1), the equal NNN hoping integrals are denoted by the same dotted lines and the unequal hopping integrals are denoted by the dotted and dashed lines, respectively.  In (a2) and (a3), the parameters are $\beta=0.5$, $V=4$ and the numerical results are $\Phi_1=\Phi_2=0.226$.  In (b2) and (b3), the parameters are $\beta=0.9$, $V=3.56$ and the numerical results are $\Phi_1=0.081$ and $\Phi_2=0.041$.  The arrows in (b2) indicate the gap broken by the SQAH order.  We use $\delta=0.04$ in the calculation of the spectral density.}
	\label{QAH_order}
\end{figure}

We use the following ansatz for the order parameters,
\begin{align}
\Phi_{AD}=\Phi_1e^{i\phi},
\quad\quad
\Phi_{BC}=\Phi_2e^{i\bar\phi},
\end{align}
where $\Phi_{1,2}$ are real.  Due to the particle-hole symmetry, $\Phi_{AD/BC}$ is purely imaginary and therefore $\phi/\bar\phi=\pm\frac{\pi}{2}$ \cite{S.W.Kim}.  Then the complex hoppings between the NNN sites can lead to the loop currents in the 2D plane.  The numerical results support two kinds of the topologically insulating QAH phases: SQAH with $\Phi_1=\Phi_2$ and AQAH with $\Phi_1\neq\Phi_2$.  The corresponding phase transitions are also continuous.  In SQAH, the loop currents connecting the NNN sites $A$ and $D$ are equal to those connecting $B$ and $C$, while in AQAH, the loop currents are unequal.  The two fermionic orders are plotted in Fig.~\ref{QAH_order}(a1) and (b1), respectively.

Within the SQAH/AQAH orders, the dispersions are given as
\begin{align}
&\varepsilon^{V_1}_{\eta\pm}(\boldsymbol k)
=\eta\Big(V^2\Phi_1^2+2(J_+^2+J_-^2)d(\boldsymbol k)
\pm 2(J_+-J_-)\sqrt{a}\Big)^\frac{1}{2}
\nonumber\\
&a=V^2\Phi_1^2 d(\boldsymbol k)+(J_++J_-)^2d^2(\boldsymbol k),
\label{V1}
\end{align}
and
\begin{align}
&\varepsilon^{V_2}_{\eta\pm}(\boldsymbol k)
=\eta\Big(\frac{V^2}{2}(\Phi_1^2+\Phi_2^2)
+2(J_+^2+J_-^2)d(\boldsymbol k)
\pm 2\sqrt{b}\Big)^\frac{1}{2}
\nonumber\\
&b=\Big(\frac{V^2}{4}(\Phi_1+\Phi_2)^2+(J_++J_-)^2d(\boldsymbol k)\Big)
\Big(\frac{V^2}{4}(\Phi_1-\Phi_2)^2
\nonumber\\
&\quad
+(J_+-J_-)^2d(\boldsymbol k)\Big),
\label{V2}
\end{align}
with $d(\boldsymbol k)$ being the same as Eq.~(\ref{dispersion}).  The dispersions for the chosen parameters are plotted in Fig.~\ref{QAH_order}(a2) and (b2).  From the dispersions, several aspects are worth pointing out: (i) For both $\varepsilon^{V_1}_{\eta\pm}$ and $\varepsilon^{V_2}_{\eta\pm}$,  although the $J_+$ and $J_-$ bands are mixed, the birefringent properties are still retained.  (ii) For the lower bands of $\varepsilon^{V_1}_{\eta-}$ and $\varepsilon^{V_2}_{\eta-}$, their minima are shifted from $k=0$ to the finite $k\neq0$.  (iii) A gap can be opened by the SQAH/AQAH order between the conduction and valence band, similar to the ACDW order.  More importantly, for the SQAH order, the two conduction (valence) bands still touch at $k=0$ that acts as the band degeneracy point, while for the AQAH order, a finite gap is opened at $k=0$ between the two conduction (valence) bands and the band degeneracy at $k=0$ is broken, as indicated by the arrows in Fig.~\ref{QAH_order}(b2).  In fact, in the region of $k\rightarrow 0$, the low-energy approximation gives: 
\begin{align}
&\varepsilon^{V_1}_{\eta\pm}(k=0)=\eta V\Phi_1,
\\
&\varepsilon^{V_2}_{\eta\pm}(k=0)=\eta V\Big( \frac{1}{2}(\Phi_1^2+\Phi_2^2)\pm\frac{1}{2}|\Phi_1^2-\Phi_2^2| \Big)^\frac{1}{2}.
\end{align}
Clearly, the gap between the two conduction (valence) bands is vanishing as $\Delta_1=0$ in the SQAH order and finite as $\Delta_2=V|\Phi_1-\Phi_2|$ in the AQAH order.

We further consider the quasiparticle spectral density $A(\omega,\boldsymbol k)$ in the momentum space, which can be experimentally measured by using the momentum resolved photoemission spectroscopy \cite{J.T.Stewart},
\begin{align}
A(\omega,\boldsymbol k)
=-\frac{1}{\pi}\text{Im}\hat G(i\omega=\omega+i\delta,\boldsymbol k),
\end{align}
here $\delta$ represents the energy spectrum broadening caused by, \textit{e.g.}, the atom linewidth and the noise-induced scatterings.  The Green's function $\hat G(\omega,\boldsymbol k)=[i\omega-\hat H_0(\boldsymbol k)+\hat H_I^d(\boldsymbol k)]^{-1}$ and $\hat H_I^d(\boldsymbol k)$ is the decoupled interaction in the mean-field framework.  In Fig.~\ref{QAH_order}(b3) of AQAH order, the finite gap between the upper/lower two bands can be clearly seen in the peaks of spectral density only if the spectrum broadening cannot smear the gap.  Thus the observation provides an important signature to distinguish the correlation-induced AQAH order from SQAH. 

In Fig.~\ref{onlyUV}(b), when $\beta<0.8$, the increasing of $V$ can drive the system from BRS into the SQAH order with the critical $V_c$ keeping unchanged, which is consistent with the previous work \cite{N.Komeilizadeh}.  While when $\beta>0.8$, the AQAH order appears, which is guaranteed by the bifurcation of the boundary line between BRS and SQAH.  In this case, the NNN interactions $V$ will first drive the birefringent fermions into the AQAH order and then into the SQAH order.  We can also see that in the limit of $\beta=1$, the lower critical $V_c$ is finite, suggesting that the depleting sublattice $D$ can make $V_c$ become smaller, but not vanishing, which is different from the behavior of the critical $U_c$ in Fig.~\ref{onlyUV}(a).  While the upper critical $V_c$ can be extended to be much large, suggesting that the AQAH order is favored by the strong NNN interactions.  To check the reliability of the AQAH, we investigate its variation with the lattice size $N_c$.  With $\beta=0.9$,  the results are plotted in the inset of Fig.~\ref{onlyUV}(b), which shows that although the finite-size effect can lead to the minor fluctuations of the phase boundaries, the AQAH order still exhibits certain stability. 

Compared with the previous work \cite{N.Komeilizadeh}, where only the SQAH order is predicted to exist, here we find that when $\beta$ is large enough, the NNN interactions tend to induce the AQAH order.  This may be attributed to the fact that in this case, $J_-\rightarrow0$, then the outer $J_-$ bands become asymptotic flatness and the linear region is much narrower (or see the DOS plot in Fig.~\ref{DOS}(d)).  In the lattice model, the sublattice $D$ will become depleted in a unit cell, leading to the AQAH orders that connect the NNN sites $A$, $D$ and $B$, $C$.  So we suggest that to see the correlation-induced effects of birefringent fermions, especially the asymptotic behavior, only adopting the low-energy dispersion may be incomplete and the whole band structure should be included.  The emergence of AQAH is definitely due to the combined effects of the interactions and the specific lattice geometry.

Some insights may be gained from the normal state susceptibilities $\chi$, as the critical strength of the interaction is inversely proportional to $\chi$, which is defined as \cite{B.Roy2017}, 
\begin{align}
\chi=-2\text{Tr}\int\frac{dk_xdk_y}{(2\pi)^2}\int\frac{d\omega}{2\pi}
[\Phi \frac{1}{i\omega-\hat H_0(\boldsymbol k)}
\Phi \frac{1}{i\omega-\hat H_0(\boldsymbol k)}],
\end{align}
with $\Phi$ being the corresponding order parameter and $H_0(\boldsymbol k)$ the noninteracting Hamiltonian.  For the QAH orders of $\Phi_{AD}$ and $\Phi_{BC}$, in the basis of $(\hat c_A,\hat c_B,\hat c_C,\hat c_D)^T$, they are given as
\begin{align}
\Phi_{AD}=i\begin{pmatrix}
&&&1
\\
&&0
\\
&0
\\
-1
\end{pmatrix},
\quad
\Phi_{BC}=i\begin{pmatrix}
&&&0
\\
&&1
\\
&-1
\\
0
\end{pmatrix}.
\end{align}
After a lengthy but straightforward calculations, the static susceptibilities are obtained in the following,
\begin{align}
\chi_{AD}=\chi_{BC}=&4\int\frac{dk_xdk_y}{(2\pi)^2}
\int\frac{d\omega}{2\pi}
\frac{\omega^2}{(\omega^2-\omega_1^2)(\omega^2-\omega_2^2)}
\nonumber\\
=&4\int\frac{dk_xdk_y}{(2\pi)^2}\frac{1}{4k}=\frac{\Lambda}{2\pi},
\end{align}
here the frequencies $\omega_1=i(1-\beta)k$, $\omega_2=i(1+\beta)k$ and $\Lambda$ is the ultraviolet cutoff of momentum in the low-energy model.  As the susceptibilities $\chi_{AD}$ and $\chi_{BC}$ are equal, so we arrive at the conclusion that both the fermionic orders that connect the sites $A$ and $D$ or $B$ and $C$ can be driven by the NNN interactions.  But their magnitudes may be equal or not, which lead to the SQAH or AQAH orders, depending on the specific parameters.

\subsection{Both Nearest-neighbor and Next-nearest-neighbor Interactions}

\begin{figure}
	\includegraphics[width=8.8cm]{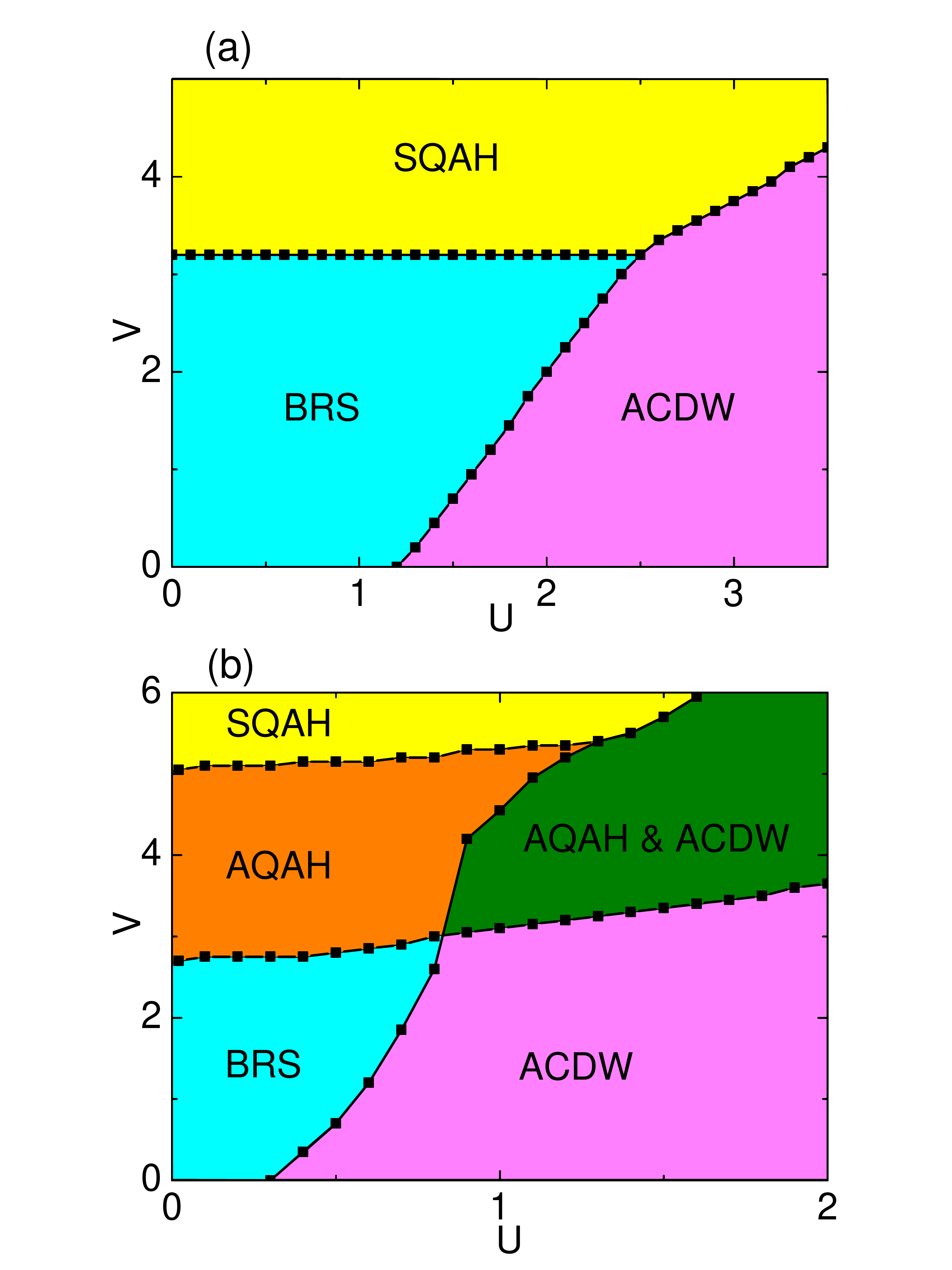}
	\caption{(Color online) Phase diagrams of the interacting birefringent fermions when both NN and NNN interactions are present, with the parameter $\beta=0.5$ in (a) and $\beta=0.9$ in (b). }
	\label{bothUV}
\end{figure}

In this section, we consider when both NN and NNN interactions are included in the system.  The interacting phase diagrams in the parametric space $(U,V)$ for $\beta=0.5$ and $0.9$ are shown in Fig.~\ref{bothUV}(a) and (b), respectively.  The phase diagrams show that the birefringent fermions are stable to the weak $U$ and $V$, \textit{i.e.}, any sufficiently weak local four-fermion interactions are irrelevant perturbations in the sense of renormalization group.  While the strong interactions can lead to the spontaneous symmetry breaking and drive the system into the final gapped Mott insulators, whose existence serves as a signature of the interacting fermion system. 

In Fig.~\ref{bothUV}(a), the phase diagram includes three phases of BRS, ACDW and SQAH.  The qualitative behavior of the phase diagram is maintained for $\beta<0.8$ \cite{N.Komeilizadeh}, although the exact positions of the boundary lines depend on $\beta$.  We observe that the NN interactions can drive the system into the CDW order with the critical interaction increasing linearly, and the NNN interactions can drive the system into the SQAH order with the critical interaction keeping unchanged.  The two boundary lines meet at the tricritical point, across which the ACDW and SQAH orders compete with each other to dominate the system.  While in Fig.~\ref{bothUV}(b), as $\beta>0.8$, the phase diagram includes five distinct phases: besides the three phases mentioned above, two additional phases interpolate in the intermediate-$V$ region which are AQAH, as expected, and the phase where the AQAH and ACDW orders coexist.  As the $J_-$ bands are not completely flat ($\beta\neq1$), here the NNN interactions $V$ will drive the system into the final SQAH orders.  The difference between the two phase diagrams in Figs.~\ref{bothUV}(a) and (b) is definitely ascribed to the asymptotic behavior of birefringent fermions lying on the specific lattice model.

The structure of the phase diagram in Fig.~\ref{bothUV}(a) is similar with other short-range interacting fermions, such as the 2D spinful Chern insulator \cite{Y.X.Wang2019}, 3D spinless hyperhoneycomb lattice \cite{S.W.Kim} and 3D spinful line-node semimetal \cite{B.Roy2017} .  Note that in the spinful system, the on-site Hubbard interactions replace the NNN interactions and consequently, the antiferromagnetic order takes the place of the QAH order.  As the phase diagram reveals the underlying mechanisms of the interaction-induced spontaneous symmetry breaking, it exhibits certain universality and we suggest that it can be extended to other fermion models subjected to the short-range interactions. 

The coexisting phase in Fig.~\ref{bothUV}(b) incorporates these two kinds of asymmetric fermionic orders and thus owns the characteristics of each order.  Such a phase occupies certain part of the phase diagram and extends to large $V$ as $U$ increases.  Evidently, its appearance is due to the correlation effects by both interactions $U$ and $V$ and requires the condition of $\beta\rightarrow1$.  The results remind us about the previous work of a 2D Weyl semimetal on a checkerboard lattice \cite{K.Sun1}.  In their work \cite{K.Sun1}, an interaction-induced phase including both the nematic and QAH orders was revealed, which is similar to our findings.  In addition, as the coexisting phase here is topologically insulating, it can also provide an insulating analog of the metallic topological nematic phase \cite{K.Sun2}.

\section{Discussions and Conclusions}
 
In a recent work about the interacting spinful birefringent fermions \cite{H.-M.Guo}, also based on the lattice model, the authors found that the sublattice magnetization and spin-spin correlation are decreasing with $\beta$ when $\beta>0.7$, and the behavior is different from the case when $\beta<0.7$.  This is also explained by the asymptotic depletion of sublattice $D$ and is consistent with the asymmetric fermion orders revealed in this work. 
 
Experimentally, the cold atoms in optical lattices provide a feasible platform to implement the interacting birefringent fermion with precise tunability and detection capability \cite{T.Esslinger}.  Specifically, the interactions between fermions are tuned via a magnetic Feshbach resonance \cite{M.Houbiers, G.Zurn}, while a scheme based on resonant modulations is developed to engineer synthetic gauge fields through the optical lattice \cite{N.Goldman}.  The ACDW order can be detected by using a band-mapping technique that maps it to the different bands of the lattice \cite{M.Schreiber,S.Trotzky} and the characteristic gap in AQAH can be captured by the momentum-resolved photoemission spectroscopy \cite{J.T.Stewart}.

To summary, we have studied the problem of the stability of interacting birefringent fermions.  Due to its specific lattice structure, the low-energy linear dispersions may be incomplete in describing the fermionic behavior.  By including the whole band structures of birefringent fermions based on the lattice model, we find the interaction-induced ACDW and AQAH orders and even the phase that incorporates both asymmetric orders.  The emergence of these asymmetric orders can broaden our understanding about the correlation effects in fermion systems.  More theoretical and experimental works about the correlated birefringent fermions are expected in the future, with the open questions including extending the present 2D model to higher dimension \cite{B.Roy2018}.

\section{Acknowledgments}

We would like to thank Linghua Wen, Biao Huang, Xiaopeng Li and W. Vincent Liu for many helpful discussions.  This work was supported by NSFC under Grant No. 11804122 (Y. X. Wang), China Scholarship Council under Grant No. 201706795026 (Y. X. Wang) and the Fundamental Research Funds for the Central Universities from China (F. Li).

\end{document}